\input{aipcheck}

\documentclass[
    ,final            % use final for the camera ready runs
  ]
  {aipproc}

\layoutstyle{6x9}

\begin{document}

\title{Central Diffraction in ALICE}


\author{R. Schicker \newline for the ALICE collaboration}{
  address={Phys. Inst., Philosophenweg 12, 69120 Heidelberg, Germany}
}

\begin{abstract}
The ALICE experiment consists of a central barrel in the 
pseudorapidity range \mbox{-0.9 < $\eta$ < 0.9} and of additional detectors
covering about 3 units of pseudorapidity on either side
of the central barrel. Such a geometry allows the  tagging
of single and double gap events. The status of the analysis
of such diffractive events in proton-proton collisions
at $\sqrt{s}$ = 7 TeV is presented. 
\end{abstract}

\maketitle

%%%%%%%%%%%%%%%%%%%%%%%%%%%%%%%%%%%%%%%%%%%%
%% MAINMATTER
%%%%%%%%%%%%%%%%%%%%%%%%%%%%%%%%%%%%%%%%%%%%

\section{Introduction}

The ALICE experiment at the Large Hadron Collider (LHC) at CERN 
consists of a central barrel covering the pseudorapidity
range -0.9 < $\eta$ < 0.9 and of a muon spectrometer
in the range -4.0 < $\eta$ < -2.4 \cite{alice}.   
Additional detectors for event classification and trigger purposes 
exist such that the range -4.0 < $\eta$ < 5.0 is covered.
The event topologies of single and double gap events can
be identified by requiring the existence of charged tracks in the 
central barrel with absence of activity on one or both sides 
of the central barrel, respectively. 

Approximately 30\% of the total proton-proton cross section
at the LHC energies is due to diffractive reaction channels, hence
detailed measurements of such channels are necessary for 
a comprehensive understanding of hadron-hadron interactions
at high energies. Moreover, central production 
is dominated by color singlet gluon exchange, hence 
such an environment offers the possibility to study 
gluonic degrees of freedom\cite{close}.

\section {ALICE detectors}

The detector systems of the ALICE central barrel track and 
identify hadrons, electrons and photons in the pseudorapidity 
range -0.9 < $\eta$ < 0.9. The magnetic field of 0.5 T 
allows the reconstruction of tracks from very low 
transverse momenta of about 100 MeV/c to fairly high values
of about 100 GeV/c. The main systems for these tasks are 
the Inner Tracking System (ITS), the Time Projection Chamber (TPC),
the Transition Radiation Detector (TRD) and the Time of Flight 
array (TOF)\cite{tpc}.

Additional detectors are placed on both sides of the 
central barrel for event classification and for trigger 
purposes. In particular, an array of scintillator detectors (V0) 
is placed on both sides and labeled V0A and V0C. These  two 
arrays cover the pseudorapidity range of
2.8 < $\eta$ < 5.1 and -3.7 < $\eta$ < -1.7, respectively.

\section{Data analysis}

The data taken with a minimum bias trigger in proton-proton 
collisions at $\sqrt s$ = 7 TeV have been analyzed according to four 
event types. Here, the mimimum bias trigger is derived
by the logical OR at L0 level from the ITS, the V0A and
V0C detectors.
Gap A and Gap C events are defined by no activity
in the V0A or V0C array, respectively. 
No gap and double gap events are defined by activity or no
activity in both V0A and V0C detectors, respectively.

The tracks are reconstructed in the central barrel,
and the characteristics of the four classes are compared.
The reconstruction of a track is subject 
to quality criteria such as number of TPC clusters > 65 and
subject to rejection of a kink topology in the track.

\begin{figure}[h]
  \includegraphics[height=.26\textheight]{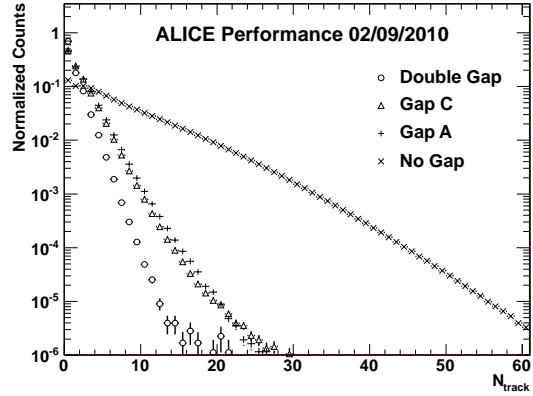}
  \caption{Central barrel multiplicity distribution of 
single, double and no gap events.}
 \label{fig1}
\end{figure}

Fig. \ref{fig1} shows the multiplicity of charged tracks 
in the central barrel for single, double and 
no gap events. These multiplicity distributions are 
normalized to unity. 

In the following, the four event classes are compared by single 
and double track observables. Exclusive production of resonances
can be selected by choosing events with two tracks in the central
barrel, hence only such events are considered in this analysis.

\begin{figure}[h]
\begin{minipage}[t]{7.2 cm}
  \includegraphics[height=.24\textheight]{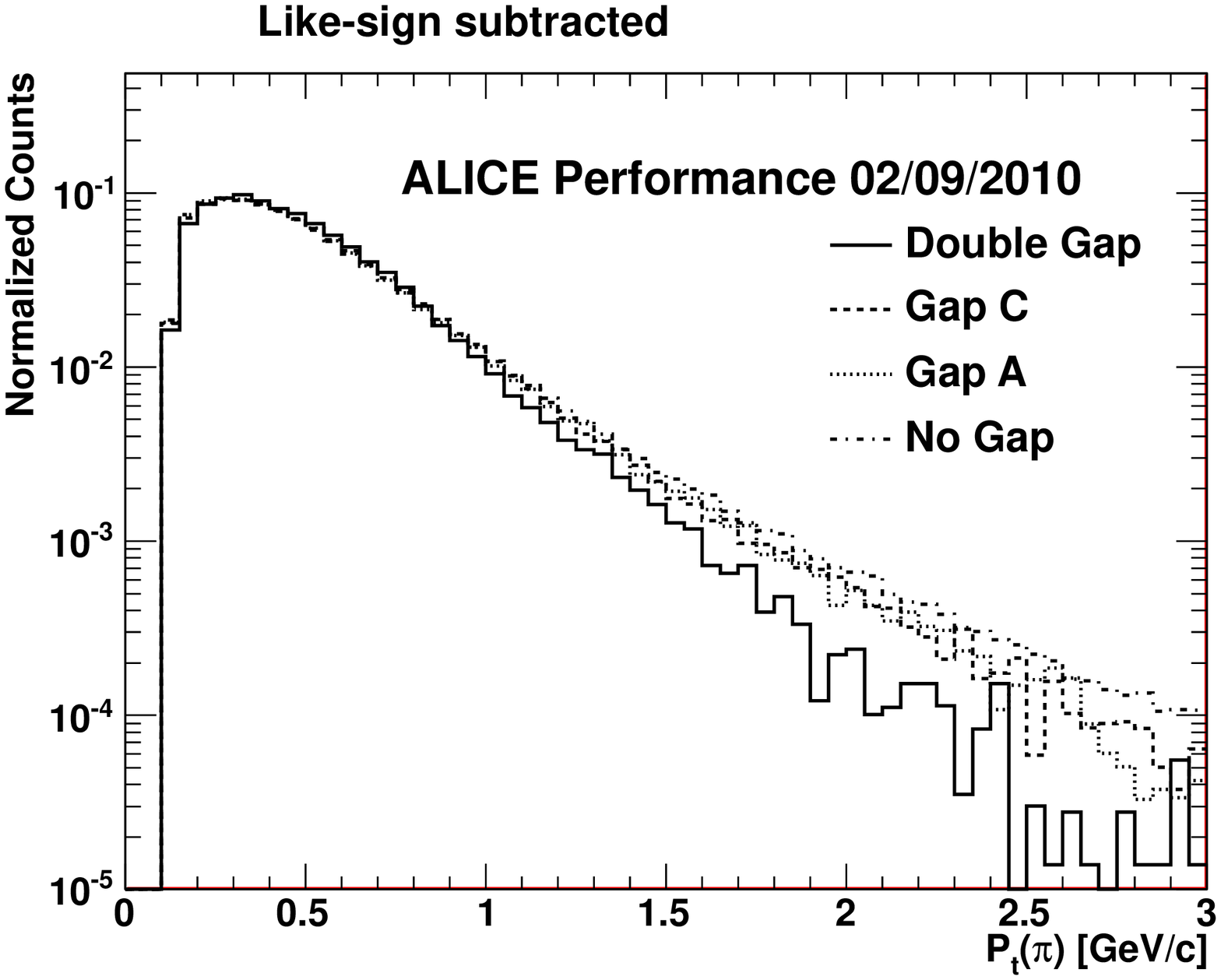}
  \caption{Single track transverse momentum distribution (left) and 
pseudorapidity distribution (right)
for  single, double and no gap events.}
 \label{fig2}
\end{minipage}
\hspace{0.0cm}
\begin{minipage}[t]{7.2 cm}
  \includegraphics[height=.24\textheight]{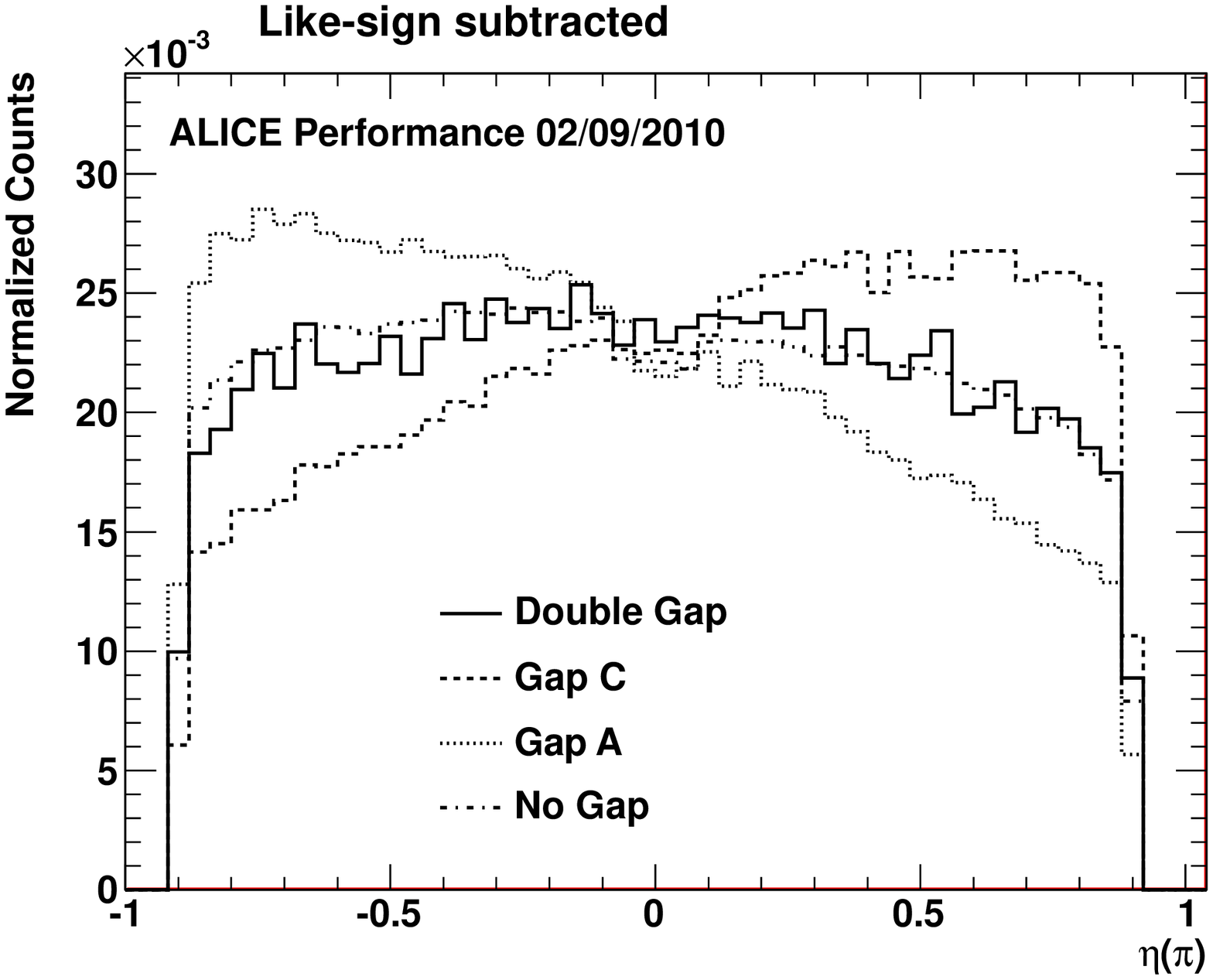}
\end{minipage}
\end{figure}

Fig. \ref{fig2} on the left shows the single track p$_T$-distribution 
for single, double and no gap events. 
Fig. \ref{fig2} on the right displays the single track pseudorapidity
distribution for single, double and no gap events. 
The distributions shown here are the raw distributions, and 
are not corrected for efficiencies of detector channels.

In this analysis, the tracks are not particle identified, and all 
tracks are assumed to be pions. The two track invariant mass 
carries the information of resonance production.

\begin{figure}[h]
\begin{minipage}[t]{7.2 cm}
  \includegraphics[height=.24\textheight]{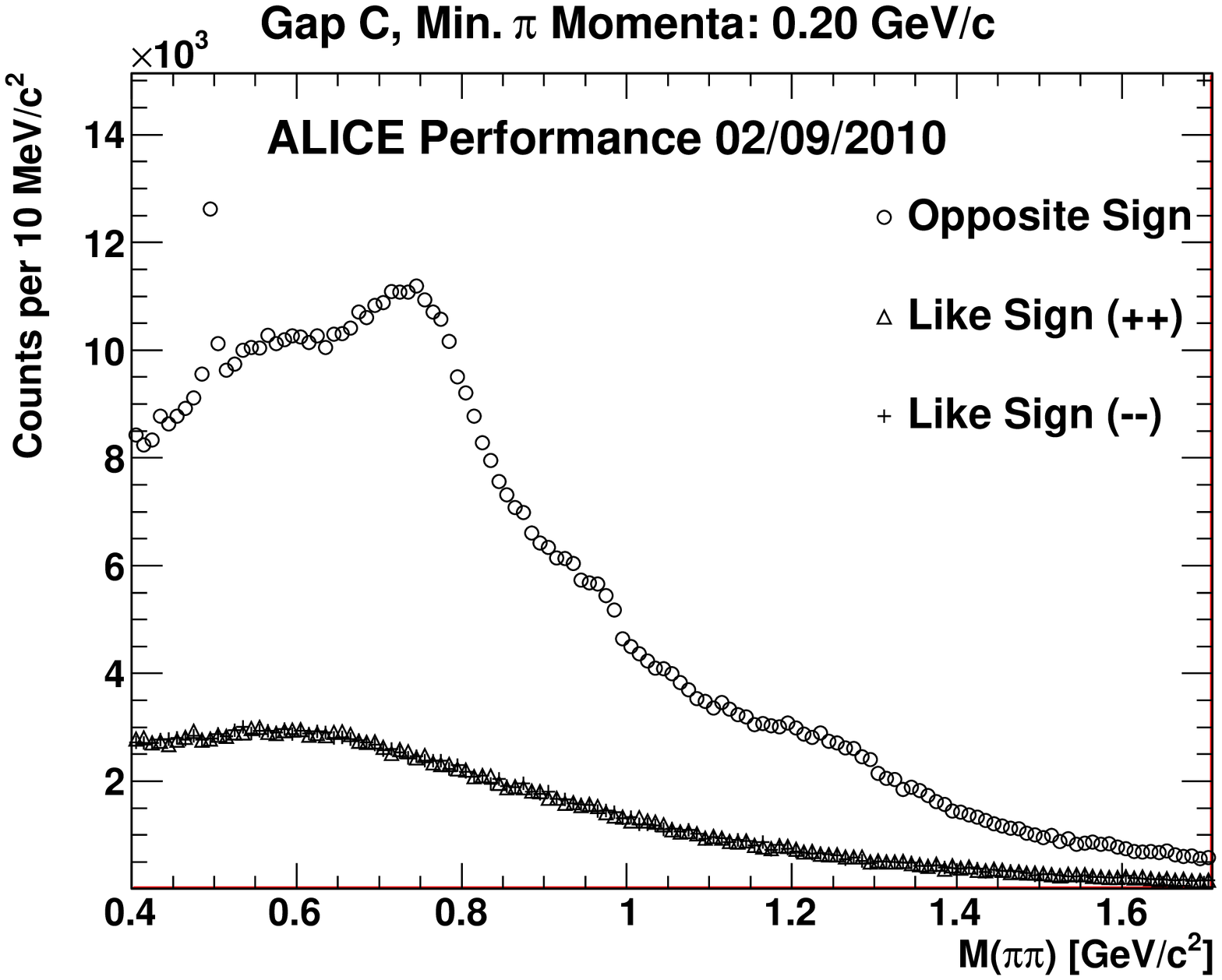}
  \caption{Two track invariant mass distribution for gap C events
with like-sign background (left) and corrected (right).}
 \label{fig3}
\end{minipage}
\hspace{0.0cm}
\begin{minipage}[t]{7.2 cm}
  \includegraphics[height=.24\textheight]{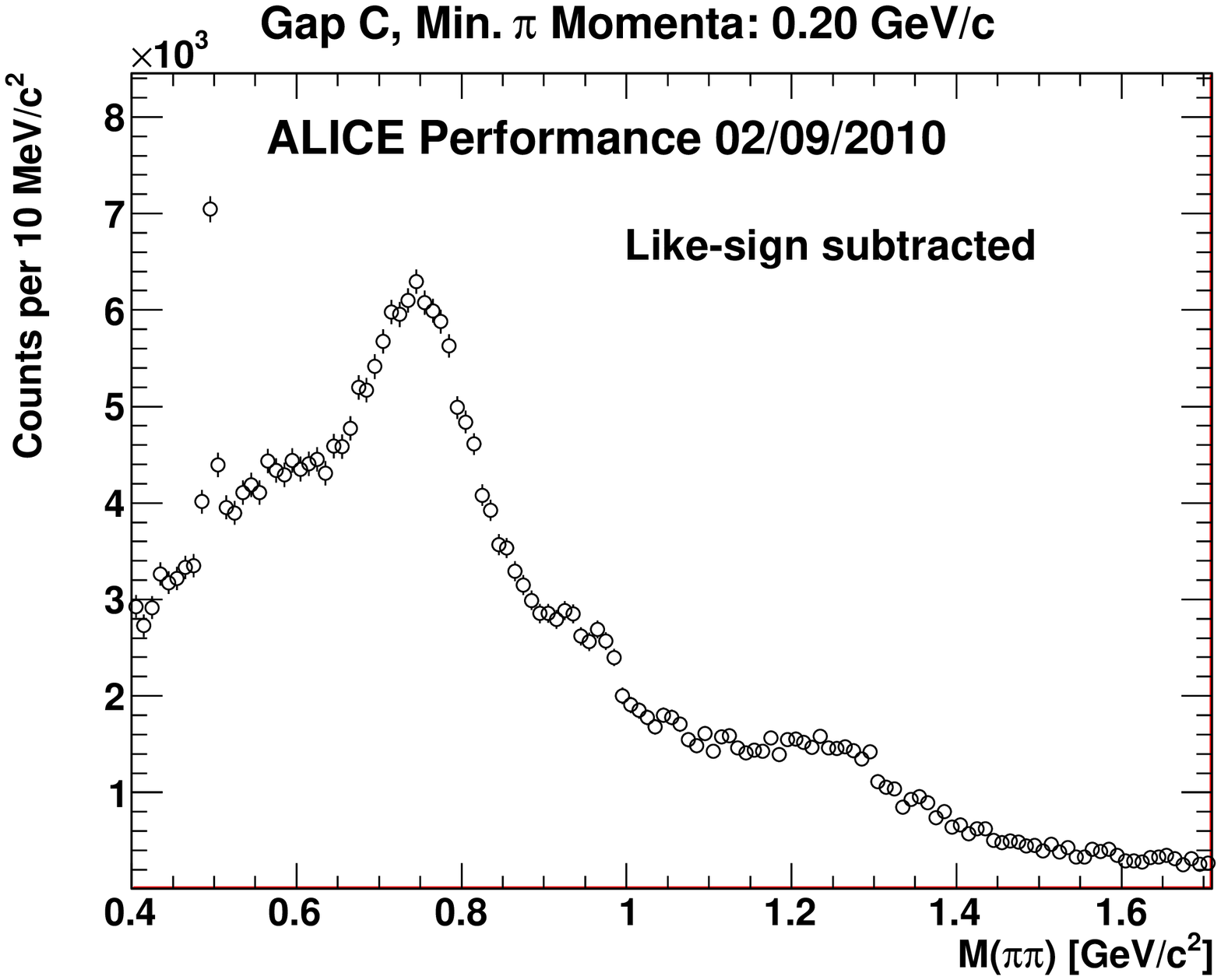}
\end{minipage}
\end{figure}
 
Fig. \ref{fig3} shows the two track invariant mass of gap C events.
Here, a single track momentum cut of 0.2 GeV/c is applied. On the 
left, the distribution of opposite-sign and like-sign pairs 
is shown. The two like-sign distributions are identical within the 
size of the data symbols reflecting a charge symmetric acceptance.
With a charge symmetric acceptance, uncorrelated tracks contribute 
equally to opposite-sign and like-sign pairs.
The uncorrelated pairs in the opposite-sign spectrum
can hence be corrected by subtracting the like-sign spectra. 
On the right, this like-sign corrected distribution is shown. 
This like-sign background
results in a correction of the raw distribution of about 40\%.

\begin{figure}[h]
\begin{minipage}[t]{7.2 cm}
  \includegraphics[height=.24\textheight]{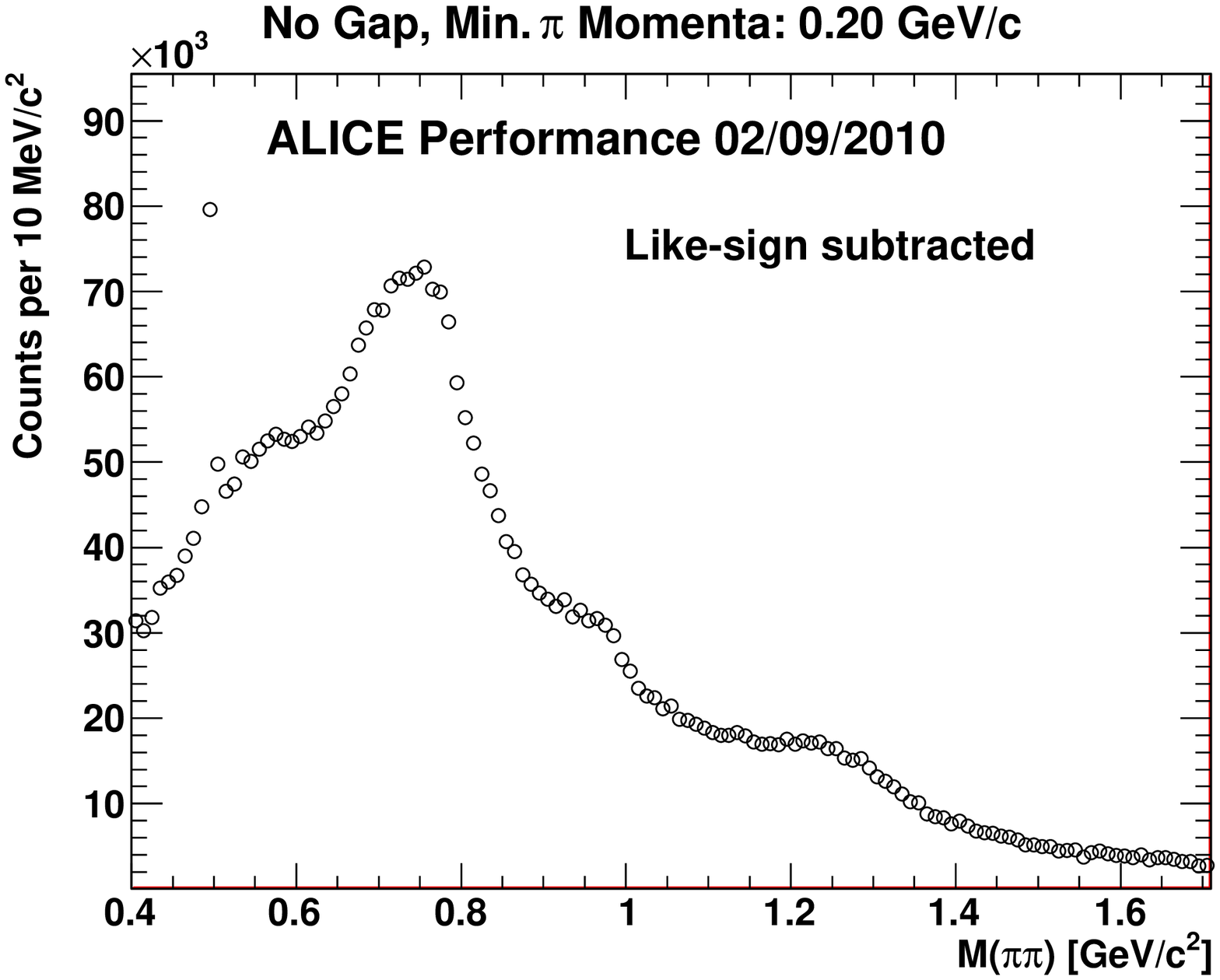}
  \caption{Two track invariant mass distribution for no  gap events
(left) and double gap events (right).}
 \label{fig4}
\end{minipage}
\hspace{0.0cm}
\begin{minipage}[t]{7.2 cm}
  \includegraphics[height=.24\textheight]{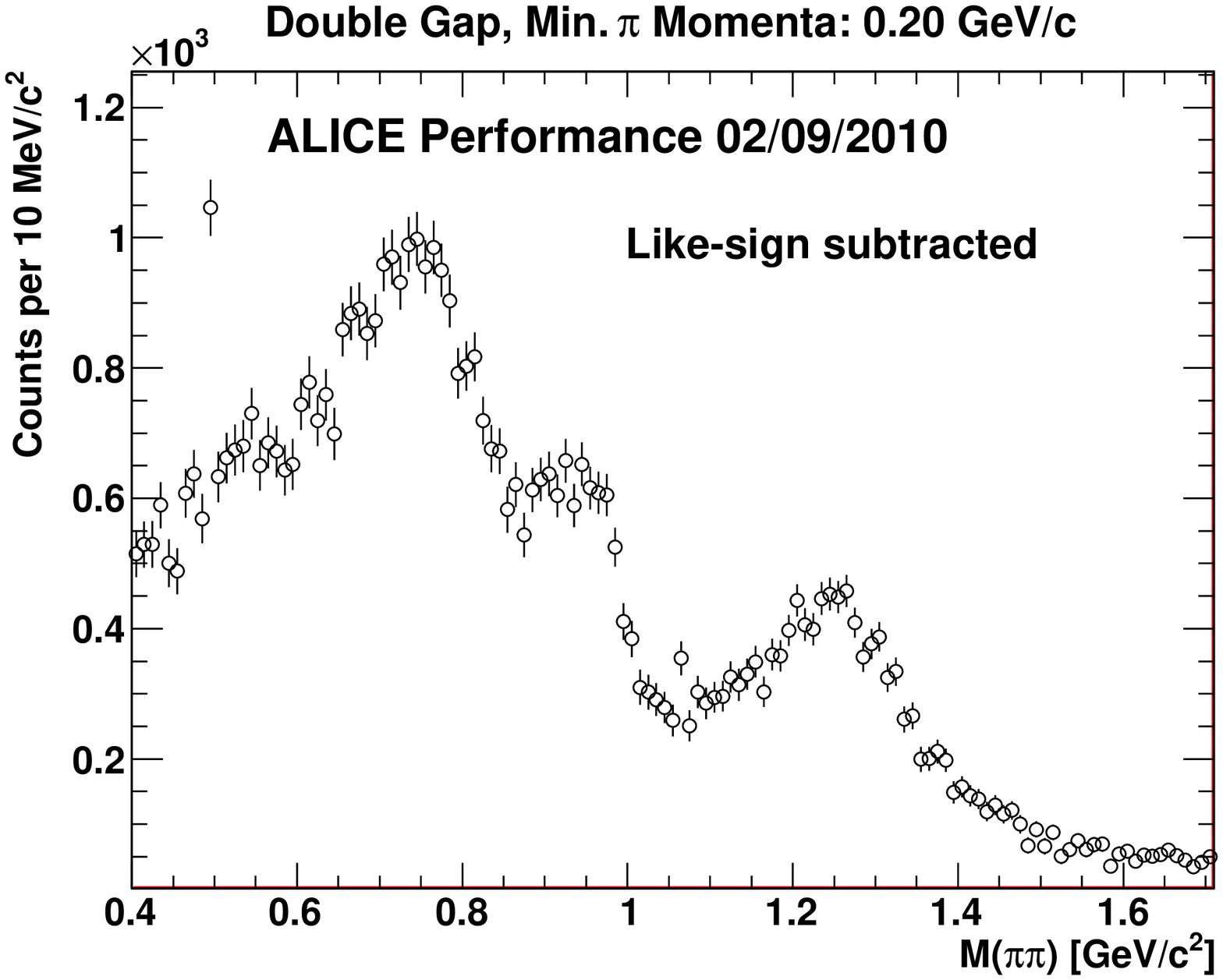}
\end{minipage}
\end{figure}

Fig. \ref{fig4} on the left displays the like-sign corrected mass 
distribution of no gap events and shows the same qualitative features as 
Fig. \ref{fig3}. In this event class, the like-sign background
results in a correction of about 60\% of the raw spectrum.  
The corrected mass distribution on the left shows a prominent 
$\rho$-signal and a K$_S^0$ signal at 0.5 GeV/c$^2$. The 
structures at the low mass tail of the $\rho$ are thought to arise
from the three body decay $\pi^{+}\pi^{-}\pi^{0}$ of the $\omega$ and 
$\eta$-meson. In these decays, only the charged tracks are seen and the 
$\pi^{0}$ escapes undetected. At the high mass tail, two 
structures are visible associated with the f$_0$(980) and 
the f$_2$(1270). 

Fig. \ref{fig4} on the right shows the corresponding
mass distribution of double gap events. Here, the like-sign background 
results in a correction of the raw spectrum of about 20\%. 

\begin{figure}[h]
  \includegraphics[height=.26\textheight]{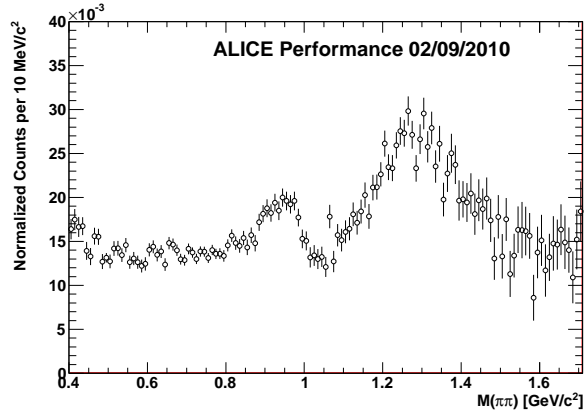}
  \caption{Mass distribution of double gap events
normalized to no gap events.}
 \label{fig5}
\end{figure}

The contribution of the f$_0$(980) and f$_2$(1270) are quantitatively 
different in the no gap and double gap events.
A normalized invariant mass distribution is defined by dividing 
the double gap mass distribution by the 
no gap mass distribution and is shown in Fig. \ref{fig5}.

Fig. \ref{fig5} shows structures
associated with an enhanced production of the resonances f$_0$(980) and 
f$_2$(1270) in double gap events as compared to no gap events.
Such an enhancement of J$^{PC}$ = J$^{++}$ (J=0,2) states
is evidence that central diffractive
production can be tagged by a double gap event topology.
The presence of J$^{PC}$= 1$^{--}$ states such as the $\rho$
in Fig. \ref{fig4} can be attributed to diffractive 
double $\rho$-production with one $\rho$ escaping detection as 
well as to non-diffractive $\rho$-production due to
detector acceptance.
The separation of the $\rho$-signal into these two sources
will be the subject of further studies of double gap events.
In particular, the non-diffractive $\rho$-component
will be reduced in the future due to an improved
pseudorapidity coverage with a new detector system \cite{ger}. 

\begin{theacknowledgments}
This work is supported in part by German BMBF under project 
06HD197D and by WP8 of the hadron physics program of the 
7$^{th}$  EU program period.
\end{theacknowledgments}



\bibliographystyle{aipproc}   % if natbib is available

%\bibliography{proc}

\begin{thebibliography}{9}

\bibitem{alice} The ALICE collaboration, K. Aamodt et al, 
The ALICE experiment at the CERN LHC, (2008) JINST\_3\_S08002

\bibitem{close}F. Close, A. Kirk, G. Schuler, Phys.Lett. B 477 (2000) 13

\bibitem{tpc}The ALICE Collaboration, J. Alme et al., Nucl. Instrum. Methods A 622 (2010) 316

\bibitem{ger} G. Herrera Corral, A new detector array for diffractive physics
in ALICE at the LHC, these proceedings


\end{thebibliography}

%\IfFileExists{\proc.bbl}{}
%  \typeout{******************************************}
%  \typeout{** Please run "bibtex \jobname" to optain}
%  \typeout{** the bibliography and then re-run LaTeX}
%  \typeout{** twice to fix the references!}
%  \typeout{******************************************}

%\endinput

\end{document}